\documentclass[twocolumn]{aastex62}

\newcommand{\he}[1] {He\,{\sc #1}}

\newcommand{\ha}{H$\alpha$}

\def\kms{\mbox{${\rm km}\:{\rm s}^{-1}\:$}}

\def\lesssim{\mathrel{\hbox{\rlap{\hbox{\lower4pt\hbox{$\sim$}}}\hbox{$<$}}}}

\def\gtrsim{\mathrel{\hbox{\rlap{\hbox{\lower4pt\hbox{$\sim$}}}\hbox{$>$}}}}

\def\vt{$v_\mathrm{t}$}

\received{\today}

\shorttitle{The optical wind of MAXI~J1820+070}
\shortauthors{Mu\~noz-Darias et al. }

\begin{document}

\title{Hard-state accretion disk winds from black holes: the revealing case of MAXI~J1820+070}


\author[0000-0002-3348-4035]{T.~Mu\~noz-Darias}
\affiliation{Instituto de Astrof\'isica de Canarias, 38205 La Laguna, Tenerife, Spain}
\affiliation{Departamento de Astrof\'\i{}sica, Universidad de La Laguna, E-38206 La Laguna, Tenerife, Spain}

\author{F. Jim\'enez-Ibarra}
\affiliation{Instituto de Astrof\'isica de Canarias, 38205 La Laguna, Tenerife, Spain}
\affiliation{Departamento de Astrof\'\i{}sica, Universidad de La Laguna, E-38206 La Laguna, Tenerife, Spain}

\author{G. Panizo-Espinar}
\affiliation{Instituto de Astrof\'isica de Canarias, 38205 La Laguna, Tenerife, Spain}
\affiliation{Departamento de Astrof\'\i{}sica, Universidad de La Laguna, E-38206 La Laguna, Tenerife, Spain}

\author{J.~Casares}
\affiliation{Instituto de Astrof\'isica de Canarias, 38205 La Laguna, Tenerife, Spain}
\affiliation{Departamento de Astrof\'\i{}sica, Universidad de La Laguna, E-38206 La Laguna, Tenerife, Spain}

\author{D. Mata S\'anchez}
\affiliation{Jodrell Bank Centre for Astrophysics, The University of Manchester, Manchester M13 9PL, UK}

\author{G. Ponti}
\affiliation{Istituto Nazionale di Astrofisica, Osservatorio Astronomico di Brera, Via E. Bianchi 46, I-23807 Merate, Italy}
\affiliation{Max Planck Institute fur Extraterrestriche Physik, D-85748 Garching, Germany} 

\author{R.~P. Fender}
\affiliation{Department of Physics, Astrophysics, University of Oxford, Keble Road, Oxford OX1 3RH, UK}


\author{D.~A.~H. Buckley}
\affiliation{South African Astronomical Observatory, PO Box 9, Observatory 7932, Cape Town, South Africa}

\author{P.~Garnavich}
\affiliation{Department of Physics, University of Notre Dame, Notre Dame, IN 46556, USA} 

\author{M.~A.~P.~Torres}
\affiliation{Instituto de Astrof\'isica de Canarias, 38205 La Laguna, Tenerife, Spain}
\affiliation{Departamento de Astrof\'\i{}sica, Universidad de La Laguna, E-38206 La Laguna, Tenerife, Spain}
\affiliation{SRON, Netherlands Institute for Space Research, Sorbonnelaan 2, NL-3584 CA Utrecht, the Netherlands}

\author{M. Armas Padilla}
\affiliation{Instituto de Astrof\'isica de Canarias, E-38205 La Laguna, Tenerife, Spain}
\affiliation{Departamento de Astrof\'\i{}sica, Universidad de La Laguna, E-38206 La Laguna, Tenerife, Spain}

\author{P.~A.~Charles}
\affiliation{Department of Physics and Astronomy, University of Southampton, Highfield, Southampton SO17 1BJ, UK} 

\author{J.~M.~Corral-Santana}
\affiliation{European Southern Observatory (ESO), Alonso de C\'ordova 3107, Vitacura, Casilla 19, Santiago, Chile}

\author{J.~J.~E. Kajava}
\affiliation{Finnish Centre for Astronomy with ESO (FINCA), FI-20014 University of Turku, Finland}

\author{E. J. Kotze}
\affiliation{South African Astronomical Observatory, PO Box 9, Observatory 7932, Cape Town, South Africa}
\affiliation{Southern African Large Telescope, P.O. Box 9, Observatory, 7935, South Africa}

\author{C.~Littlefield} 
\affiliation{Department of Physics, University of Notre Dame, Notre Dame, IN 46556, USA} 

\author{J.~S\'anchez-Sierras}
\affiliation{Departamento de Astrof\'\i{}sica, Universidad de La Laguna, E-38206 La Laguna, Tenerife, Spain}

\author{D.~Steeghs}
\affiliation{Department of Physics, University of Warwick, Gibbet Hill Road, Coventry CV4 7AL, UK}

\author{J. Thomas}
\affiliation{South African Astronomical Observatory, PO Box 9, Observatory 7932, Cape Town, South Africa}

\begin{abstract}

We report on a detailed optical spectroscopic follow-up of the black hole transient MAXI~J1820+070 (ASASSN-18ey). The observations cover the main part of the X-ray binary outburst, when the source alternated between hard and soft states following the classical pattern widely seen in other systems. We focus the analysis on the \he{i} emission lines at 5876 and 6678 \AA , as well as on \ha .  We detect clear accretion disk wind features (P-Cyg profiles and broad emission line wings) in the hard state, both during outburst rise and decay. These are not witnessed during the several months long soft state. However, our data suggest that the visibility of the outflow might be significantly affected by the ionisation state of the accretion disk.  The terminal velocity of the wind is above $\sim 1200$ \kms, which is similar to outflow velocities derived from (hard-state) optical winds and (soft-state) X-ray winds in other systems. The wind signatures, in particular the P-Cyg profiles, are very shallow, and their detection has only been possible thanks to a combination of source brightness and intense monitoring at very high signal-to-noise. This study indicates that cold, optical winds are most likely a common feature of black hole accretion, and therefore, that wind-like outflows are a general mechanism of  mass and angular momentum removal operating throughout the entire X-ray binary outburst. 

\end{abstract}


\defcitealias{Munoz-Darias2016}{MD16} 
\section{Introduction}

Disk winds are observed in accreting black holes (BHs) across the full range of masses \citep{Fabian2012, Ponti2016, DiazTrigo2016}. In stellar-mass BHs, X-ray winds have been established as a fundamental property of their most radiatively efficient outburst phases, the so-called soft states, likely impacting on the accretion process. However, these highly ionised (hot) winds are scarcely observed during the typically dimmer hard states, where most of the BHs of the Universe exist and kinetic feedback from jets dominates \citep{Miller2006,Neilsen2009, Neilsen2011, Ponti2012,Fender2016}. The disappearance of the wind in the hard state is a matter of strong debate and has been suggested to be related to different physical processes, such as the details of the wind-launching mechanism, photoionization instabilities and over-ionization of the ejecta \citep[e.g.][]{Chakravorty2013,Bianchi2017,Gatuzz2019}. In addition, the presence of a continuous, state-independent wind has been proposed as a viable mechanism to explain the high efficiency of angular momentum removal inferred from fits to X-ray light curves \citep{Tetarenko2018}. Low-ionisation (cold) disk winds, on the other hand, have been detected via optical/infrared observations in the luminous and violent outbursts of V404 Cyg [\citealt{Munoz-Darias2016} (hereafter \citetalias{Munoz-Darias2016}), \citealt{Munoz-Darias2017,Rahoui2017,MataSanchez2018}; see also \citealt{Casares1991}] and V4641 Sgr \citep{Chaty2003, Lindstrom2005, Munoz-Darias2018}. These are also thought to severely impact on accretion properties, with mass outflow rates significantly exceeding the accretion rate (\citetalias{Munoz-Darias2016}, \citealt{Casares2019}). 

MAXI~J1820+070 (hereafter J1820+070) was initially discovered as an optical transient by the All-Sky Automated Survey for Supernovae (and named ASASSN-18ey; \citealt{Tucker2018}) and then as an X-ray transient  by the Monitor of All-sky X-ray Image (MAXI; \citealt{Matsuoka2009}). It was classified as a BH candidate based on its multi-wavelength properties \citep{Kawamuro2018, Kennea2018, Baglio2018, Bright2018, Shidatsu2018, Tucker2018}. The system is among the brightest BH transients ever observed \citep{Corral-Santana2016}, with an X-ray flux of $\sim$ 4 Crab and an optical magnitude of $g\sim$ 11.2 at outburst peak \citep{Shidatsu2019}. In this letter, we report on an intensive optical spectroscopic follow-up campaign obtained with a suite of the largest telescopes providing high signal-to-noise data.

\section{Observations}
We carried out spectroscopy during 37 epochs, which cover in great detail the different phases of the outburst (see Table \ref{log}).

\begin{table}
\caption{Observing log, X-ray state and emission line properties.}

\begin{tabular}{clclclclclc|}
Epoch & Date (hh:mm) & Tel. & State\tablenotemark{$^{a}$} & He~\sc{i}\tablenotemark{$^{b}$}  & \ha \tablenotemark{$^{b}$} \\

\hline
1 & 15/03 (14:46) & Keck &  Hard&  -- & P-Cyg(*)\\
2 & 16/03 (08:03) & VLT & Hard & P-Cyg  & P-Cyg (*)\\  
3 & 17/03 (05:26) & GTC & Hard & P-Cyg & P-Cyg(*)\\  
4 & 18/03 (06:09)  & GTC & Hard & BW & BW \\  
5 & 20/03 (05:58) & GTC & Hard & BW &BW \\
6 & 20/03 (08:07) & VLT & Hard & BW & BW \\  
7 & 21/03 (06:11) & GTC & Hard & P-Cyg & BW+P-Cyg \\
8 & 22/03 (05:38)  & GTC & Hard & & BW \\  
 9 & 22/03 (07:53) & VLT & Hard & & BW\\  
10 & 24/03 (05:38) & GTC & Hard & & BW\\
11 & 26/03 (04:29) & GTC & Hard & P-Cyg & BW\\
12 & 23/04 (02:26) & SALT & Hard & P-Cyg(*) & BW+P-Cyg\\
13 & 13/05 (02:03)	& SALT & Hard &  & \\
14 & 14/05 (01:04) & SALT  & Hard & &\\
15 & 17/05 (00:53) & SALT & Hard & &\\
16 & 17/06 (21:26) & TNG & Hard & -- & \\
17 & 18/06 (04:55) & TNG & Hard & -- & \\
18 & 18/06 (21:59) & TNG & Hard & -- & \\
19 & 08/07 (02:05) & GTC & Soft & ?  &\\
20 & 10/07 (21:33) & GTC & Soft & &\\
21 & 11/07 (21:26) & GTC & Soft & &\\
22 & 13/07 (01:29) & VLT & Soft & &\\  
23 & 15/07 (01:28) & GTC & Soft & &\\
24 & 18/07 (01:42) & GTC & Soft & &\\
25 & 18/07 (21:42) & GTC & Soft & &\\
26 & 24/07 (22:49) & GTC & Soft & &\\
27 & 27/07 (21:29) & GTC & Soft & &\\
28 & 03/08 (23:20) & GTC & Soft & &\\
29 & 09/08 (22:18) & GTC & Soft & &\\
30 & 15/08 (21:41) & GTC & Soft & &\\
31 & 19/08 (21:56) & GTC & Soft & &\\
32 & 07/09 (00:58) & VLT & Soft & &\\  
33 & 28/09 (23:57) & VLT & Hard & P-Cyg & BW \\   
34 & 29/09 (23:51) & VLT & Hard & P-Cyg &\\  
35 & 12/10 (21:35) & GTC & Hard & & \\
36 & 21/10 (21:19) & GTC & Hard & & \\
37 & 04/11 (19:45) & GTC & Hard & & \\ 
\hline
\end{tabular}
\tablenotetext{a}{X-ray state based on MAXI data}
\tablenotetext{b}{Detection of P-Cyg profiles and broad emission line wings (BW). See text for those marked (*).  The hyphen (--) indicates lack of spectral coverage.}

\label{log}
\end{table}

\subsection{Gran Telescopio Canarias (GTC)}
We observed the target with OSIRIS (\citealt{Cepa2000}) attached to the GTC, at the Observatorio del Roque de los Muchachos (ORM) in La Palma, Spain. We used grisms R2500R (5575 -- 7685 \AA) and R2500V (4500 -- 6000 \AA), which combined with a 1.0 arcsec slit provide a velocity resolution of $\sim$  200 \kms. In total we obtained 93 spectra across 22 different nights with on-source times in the range of 300--840~s.

\subsection{Very Large Telescope (VLT)} 
J1820+070 was observed in 7 different epochs with the X-shooter spectrograph \citep{Vernet2011} attached to the VLT-UT2 at Cerro Paranal, Chile. The instrument has 3 different arms covering the ultraviolet, visible and near-infrared spectral ranges, but only the first two were included in this study (the full spectroscopic database will be presented in a forthcoming work). For each epoch, we obtained 8--16 exposures with total on-source time in the range of 1560--2256 s. We used slit-widths of 0.9 and 1.0 arcsec in the visible and ultraviolet arms, which rendered a velocity resolution of $\sim  35$ \kms.

\subsection{The Southern African Large Telescope (SALT)}
Spectroscopy was obtained on 4 separate nights with SALT (\citealt{Buckley2006}) situated at the South African Astronomical Observatory in Sutherland, using the Robert Stobie Spectrograph (\citealt{Burgh2003}). We used the PG1300 grating and a 1.5 arcsec slit-width, which gave a wavelength coverage of 4580$-$6630 \AA\ at a mean velocity resolution of $\sim  200$ \kms. Sets of continuous 25~s exposures were obtained and combined (on-source time between 1950 and 2450~s).

\subsection{Telescopio Nazionale Galileo (TNG)}
We observed the target with the TNG during 3 epochs over 2 consecutive nights. Observations were taken with the spectrograph DOLORES. We used  a slit-width of 1.0 arcsec together with grisms V486 (4612-- 4838 \AA; 400~s on-source) and VHR-R (6238--7717 \AA; 300~s on-source), which resulted in velocity resolutions of 50 and 120 km s$^{-1}$, respectively. 

\subsection{Keck Observatory} 
Our first observing epoch was taken with the LRIS \citep[][]{Oke1995} attached to the Keck-I telescope on Maunakea (Hawaii, USA). The red and blue arms of the spectrograph were divided using the 560~nm dichroic. On the blue side we employed the 600/4000 grism (3040--5630 \AA), while on the red arm we used the 900/5500 grating (6034--8440 \AA). The 0.7~arcsec wide slit provided a velocity resolution in the range of 100--180 \kms.  A total of $12\times15$ s spectra were obtained in the red and $16\times15$ s spectra in the blue.

\begin{figure}
\epsscale{1.20}
\plotone{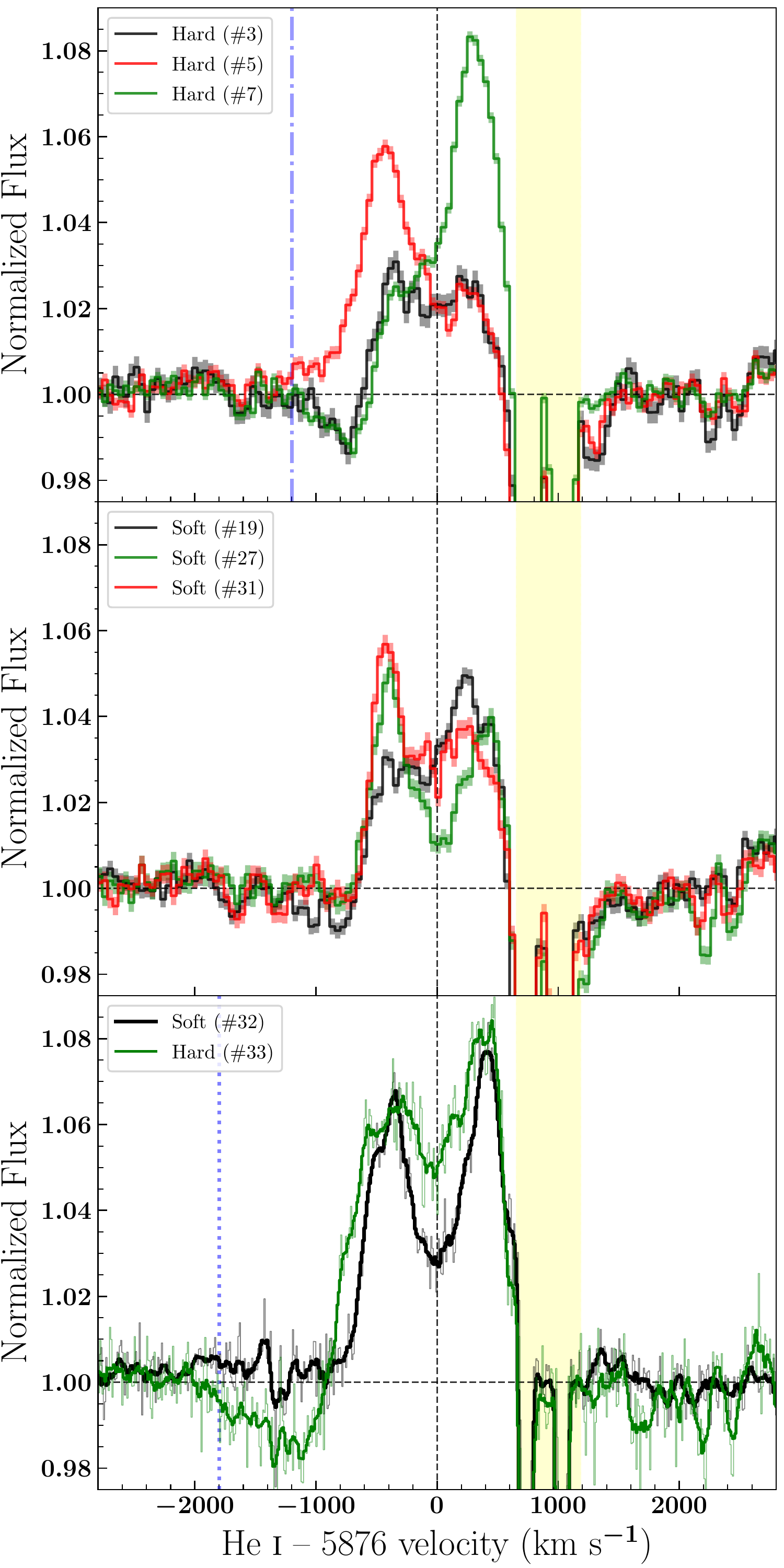}
\caption{Evolution of \he{i}--5876. \textit{Top panel:} example of hard-state wind features detected during outburst rise. P-Cyg profiles show the presence of a wind with \vt $\sim 1200$ \kms (dash-dotted line). One epoch showing a broad (blue) wing is also represented (red spectrum). \textit{Middle panel:} examples of soft state observations where no significant wind features are found. For epoch \#19 the blue edge might be compatible with a P-Cyg feature (see text). \textit{Bottom panel:} P-Cyg profiles are absent in the last soft state observation (\#32) but they re-appear over the soft-to-hard transition (epoch \#33) with \vt$\sim 2000$ \kms (dotted line). Yellow shading indicates regions contaminated by interstellar absorption.
\label{fig:pcyg}}
\end{figure}

\section{Analysis and Results}
The long-slit spectra were reduced, extracted, wavelength and flux calibrated using standard IRAF tools or the PySALT package \citep{Crawford2016} tailored for SALT data. The X-shooter spectra were processed and combined using version 3.2.0 of the EsoReflex pipeline. We used {\sc molly} and custom software under {\sc python} to perform the analysis. In order to increase the signal-to-noise ratio we henceforth focus on average spectra computed for each observing epoch (Table \ref{log}). 

\begin{figure}
\epsscale{1.15}
\plotone{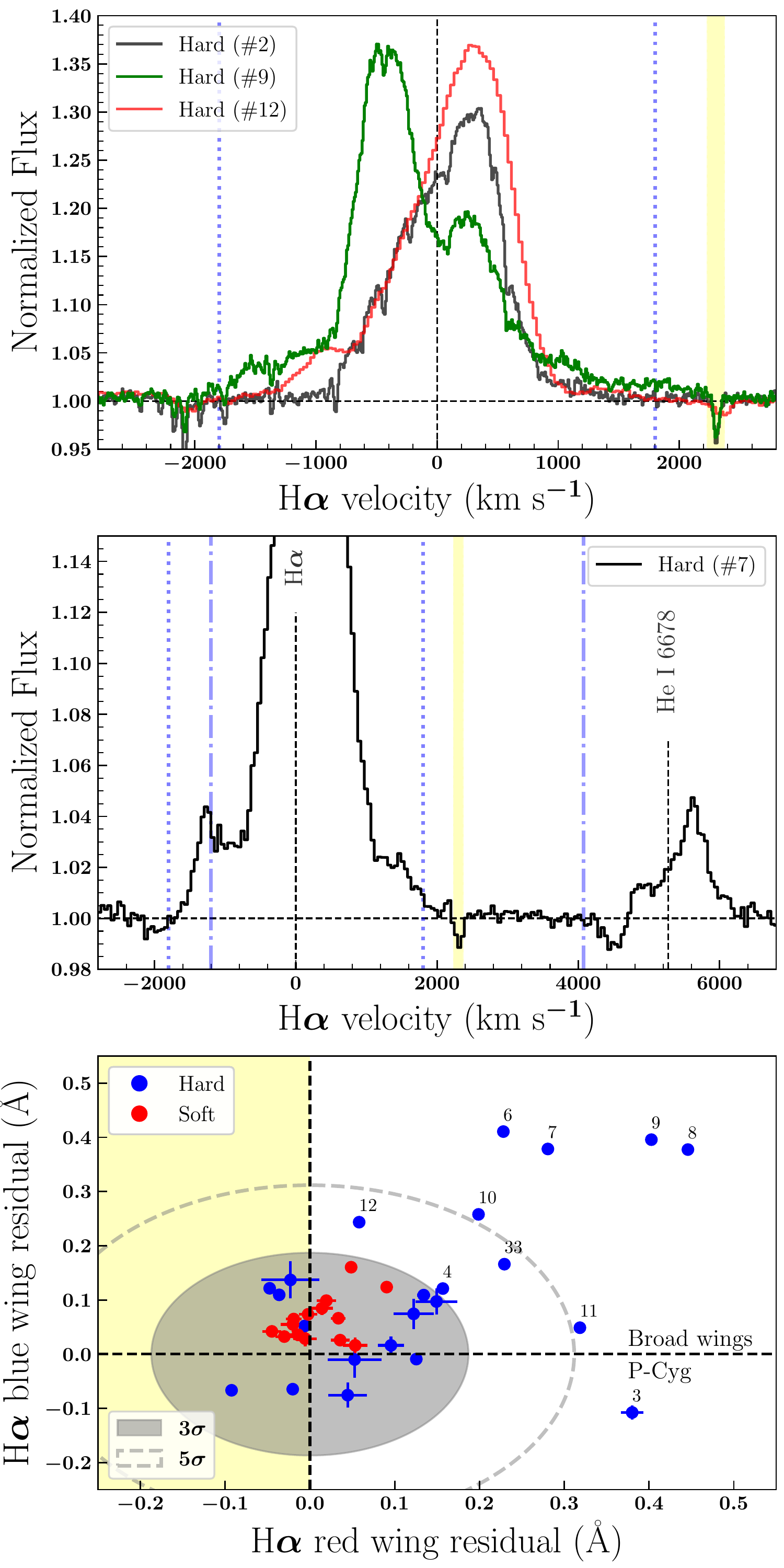}
\caption{Evolution of \ha. \textit{Top panel} shows the best examples for asymmetric lines (black, red) and broad emission line wings (green). \textit{Middle panel} displays epoch \#7, where P-Cyg profiles (\he{i}-6678) and broad wings (\ha) are detected. \ha\ also shows an absorption trough  with a blue edge velocity of $\sim -1200$ \kms. The two \vt\ inferred from P-Cyg profiles (\he{i}-5876) at different stages of the outburst (dotted and dash-dotted lines at $\pm$ 1800 and -1200 \kms) align very closely with the velocities of the \ha\ wind features. The dash-dotted line on the right corresponds to -1200 \kms with respect to \he{i}-6678.  Yellow shading indicates regions contaminated by interstellar absorption. \textit{Bottom panel:} \ha\ wings diagram. Epochs (note associated numbers) beyond the $5\sigma$  contour (dashed line) show conspicuous emission line wings. None of the soft-state data points (red dots)  is located beyond the $3\sigma$ gray-shadowed area.  The left regions of the diagram are forbidden for typical wind features (yellow shading). }
\label{fig:ha}
\end{figure}

As previously reported by \citet{Tucker2018},  the spectra are very rich in emission lines, including the Balmer series as well as several \he{i}  and \he{ii} transitions. Our multi-epoch observations show that their shape and strength change dramatically throughout the outburst. We observe strong line asymmetries, especially during the early stages of the outburst, as well as shallow P-Cyg profiles and broad, non-Gaussian emission line wings. These features are standard accretion disk wind signatures, similar to those observed in V404 Cyg (\citetalias{Munoz-Darias2016}) and V4641 Sgr \citep{Munoz-Darias2018}. Figs. \ref{fig:pcyg} and \ref{fig:ha} show a suite of the different emission line behaviours present in our data, albeit we do not show every spectrum for clarity reasons (however, see Table \ref{log} for a full summary). In particular, P-Cyg profiles are best detected in  \he{i}~5876 \AA\  (\he{i}--5876  hereafter), whilst asymmetries and broad emission wings are best seen in \ha, which is the strongest optical emission line.  We note that these are also typical outflow tracers in other stellar classes.  In particular, \he{i}--5876 is one of the best optical wind markers in massive stars, and P-Cyg profiles in this line together with broad \ha\ wings are a common feature in O supergiants \citep[e.g.][]{Prinja1994}. Likewise, disk-dominated accreting white dwarfs can show, besides P-Cyg profiles in ultraviolet resonance lines, wind signatures in \he{i}--5876 and \ha\ \citep{Kafka2004}.

We focussed the analysis on three narrow spectral ranges covering the following lines: (i) \ha\ (6563~\AA) and the adjacent \he{i}~6678 \AA, (ii) \he{i}--5876, and (iii) a third region covering the Bowen blend (mainly N~\textsc{iii} at 4641 \AA) and \he{ii}~4686 \AA\ (B+\he{ii} hereafter). We have searched for winds in the former two regions, while the last one was used as a proxy for the ionization state of the disk.  The three bands are covered by the whole dataset with the exception of \he{i}--5876, which is absent in the (four) Keck and TNG epochs. Every spectral region was carefully normalized in each of the 37 average spectra by fitting the adjacent continuum with a first order polynomial. This resulted in accurate continuum normalizations that were individually inspected.  

\begin{figure}[t]
\epsscale{1.15}
\plotone{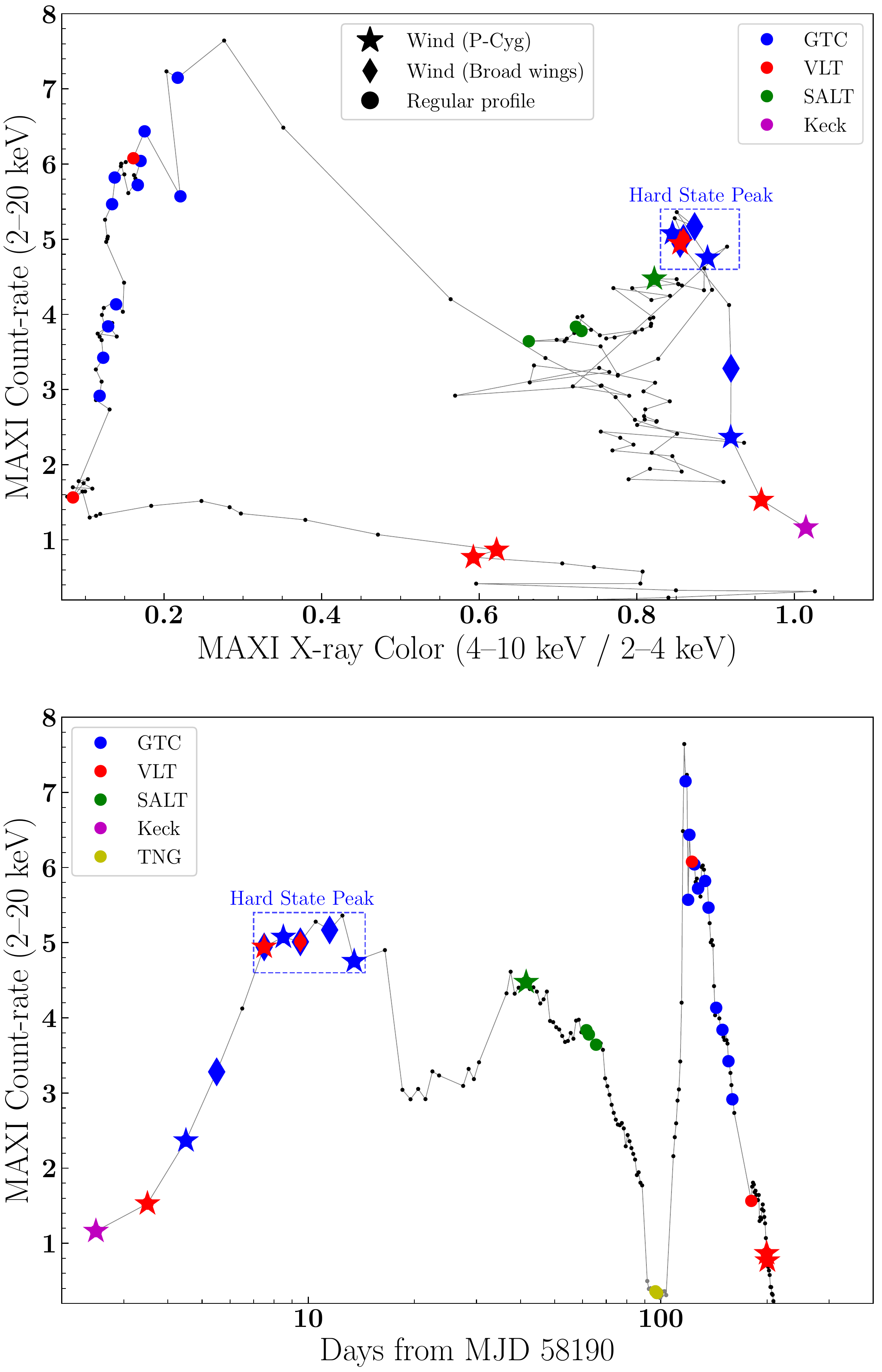}
\caption{Wind detections as a function of the X-ray outburst evolution. \textit{Top panel:} hardness-intensity-diagram from the MAXI (black dots; daily average light-curve). Simultaneous (within 0.5 d) optical observations are marked according to wind signatures and telescope (see legend). \textit{Bottom panel}: corresponding MAXI X-ray lightcurve. Grey points at very low flux (day $\sim 100$) are corrupted and only included to indicate the TNG epoch times. The hard state peak, when the most conspicuous wind signatures are witnessed, is indicated by blue rectangles.
\label{fig:hid}}
\end{figure}

In addition, we used data from MAXI to infer the X-ray state at the time of each epoch (see Table \ref{log}). We built the hardness-intensity-diagram \citep{Homan2001} using MAXI standard bands and daily averaged fluxes (Fig. \ref{fig:hid}). As was shown by \citet{Shidatsu2019}, the system displayed the q-shaped diagram typically observed in BH transients (e.g. \citealt{Mcclintock2006,Belloni2011} for reviews). Our spectroscopic observations cover in great detail the evolution of the system throughout both the hard and soft X-ray states.  We define hard and soft spectroscopic epochs as those with X-ray colors\footnote{Count-rate ratio between 4--10 keV and 2--4 keV.} higher and lower than $0.4$, respectively.  As a proxy for the luminosity (i.e. intensity) we used the 2--20 keV MAXI count-rate, but note that a significant amount of the total flux is expected to be below 2 keV during the soft state. 

\subsection{P-Cygni profiles}
The detection of a P-Cyg profile, first observed in Wolf-Rayet stars \citep{Beals1929}, is conclusive evidence for the presence of an outflow. In compact binaries, however, strong emission line components are naturally produced in their accretion disks \citep{Smak1969}, leaving the blue-shifted absorption of the P-Cyg profile as the only unaltered and unambiguous wind signature.  Therefore, we have searched for blue-shifted absorptions in all our 37 epochs. We initially focussed on \he{i}--5876 and subsequently extended the search to the \ha\ region (see Table \ref{log} for a summary).\par

Six epochs exhibit broad \he{i}--5876 blue-shifted absorptions reaching flux values below 99 per cent of the continuum level (i.e. depth $> 1$ per cent; top panel in Fig. \ref{fig:pcyg}). Given the quality of our data, we estimate that this is a secure threshold to infer the presence of P-Cyg absorption. \ha\ also shows wind-related features in five of the six epochs (see below). We will refer to the velocity of the blue edge of the P-Cyg absorption component as the wind terminal velocity (\vt). Following \citetalias{Munoz-Darias2016}, we estimate \vt\ by fitting a two Gaussian model to the data (one in absorption and one in emission) with \vt\ corresponding to the velocity at 0.1 of the maximum depth of the (fitted) absorption. We find that, although the absorptions are not typically fully Gaussian, this method provides a good description of the depth of the profile and the velocity of the blue edge. During the outburst rise, we measure \vt = $1206\pm13$ \kms when fitting simultaneously the three GTC observations ($\#3$, $\#7$ and $\#11$), while the individual fits (that includes $\#2$, taken with the VLT) provide consistent (but less constraining) results. Therefore, we take \vt $\sim 1200$ \kms as the wind velocity during this stage (dash-dotted, vertical lines in Fig. \ref{fig:pcyg}). We note that blue-shifted absorptions with the same \vt\ are sometimes detected in \ha\ and the weaker \he{i}~6678 \AA~(middle panel in Fig. \ref{fig:ha}).  Likewise, we measure  \vt = $1820\pm60$ \kms by fitting observations \#33 and \#34, and take \vt $\sim 1800$ \kms as the wind velocity during the outburst decay (dotted vertical lines in Figs. \ref{fig:pcyg} and \ref{fig:ha}). This velocity is significantly larger than that inferred from P-Cyg absorptions during the outburst rise. 

Unfortunately, we do not have \he{i}--5876 coverage during epoch \#1. However, \ha\ is very asymmetric towards the red (not shown) and by comparing it with epoch $\#2$ (Fig. \ref{fig:ha}, top panel)  taken only 17~hrs apart, we notice that the latter is just an evolution (towards more asymmetry) of the former.  On the other hand, epoch $\#12$ shows a shallow absorption in \he{i}--5876 ($\sim1$ per cent), while \ha\ is asymmetric, has broad emission line wings and an absorption trough compatible with \vt $\sim 1000$--$1200$ \kms (Fig. \ref{fig:ha}, top panel). Including these two epochs ($\#1$ and $\#12$), we conclude that eight epochs showed P-Cyg features. 

Finally, three consecutive epochs ($\#4$, $\#5$ and $\#6$) show a broad emission line wing in \he{i}--5876 (red solid line Fig. \ref{fig:pcyg} top panel) reaching $\sim 1200$ \kms, hence very similar to \vt\ measured from P-Cyg absorptions before and after these epochs. This behaviour has been also witnessed in V404 Cyg (\citealt{MataSanchez2018}) and we interpret these features as wind detections. The remaining epochs, including every soft state observation (Fig. \ref{fig:pcyg} middle panel), do not show wind signatures in \he{i}--5876 (nor in \ha; see below). A possible exception might be  \#19, the first soft state observation, which shows relatively complex and weak blue-shifted absorptions (middle panel in Fig. \ref{fig:pcyg}). Also, we do not detect P-Cyg absorptions during the last 3 epochs, when the system decayed through the faint hard state below the MAXI detection limit.

\subsection{The \ha\ spectral region}
Several epochs show broad \ha\ wings, as well as strong line asymmetries (top panel in Fig. \ref{fig:ha}). Interestingly, the wings extend up to velocities remarkably consistent with those derived from the P-Cyg profiles observed months later, during the decay of the outburst (i.e. \vt $\sim 1800$ \kms; dotted, vertical lines in Figs. \ref{fig:pcyg} and \ref{fig:ha}). Hence, they exceed \vt $\sim 1200$ \kms measured  from \he{i}--5876 in the very same observations. This behaviour is clearly exemplified by epoch $\#7$ (hard state peak). It shows, in addition to a P-Cyg profile in \he{i}--5876 (\vt $\sim 1200$ \kms ; Fig. \ref{fig:pcyg}), strong emission line wings in \ha\ (reaching $\sim 1800$ \kms) and a superimposed absorption trough with a blue edge velocity of -1200  \kms.  A (regular) P-Cyg profile in \he{i}~6678 \AA\ at \vt $\sim 1200$ \kms is also present (middle panel in Fig. \ref{fig:ha}). 

The strength of \ha\  and our data's high signal-to-noise allow for a systematic search of wind features using the diagnostic diagram developed in  \citet{MataSanchez2018} for V404~Cyg.  To this end, we performed a Gaussian fit to the line profile that was subsequently subtracted from the data. We masked the innermost part of the line (-300 to 300 \kms in velocity scale), which can be affected by the double-peak and other line core asymmetries, as well as the emission line wings that we are trying to detect ($\pm 1000$ to  $\pm 2000$ \kms). In the bottom panel of Fig. \ref{fig:ha} we plot the equivalent width (EW) of residuals in the blue (-1800 to -1000 \kms) and red (1000 to 1800 \kms) emission line wings\footnote{The masks were shifted by $\pm 200$ \kms for the last 5 epochs to account for the increasing breadth of the line during the decay.} . Significance levels are computed by measuring the \textit{EW of the continuum}  within masks of the same width (i.e. 800 \kms) in nearby continuum regions. These \textit{continuum residuals} show a Gaussian-like distribution (with mean equal to $\sim 0$) when considering the full sample of spectra.  This is used to trace significance contours in Fig. \ref{fig:ha}. The 9 epochs with residuals exceeding the $3\sigma$ level are hereafter quoted as wind detections (Table \ref{log}). They all correspond to hard state epochs.  We note that both the Gaussian fits and the associated residuals were individually inspected, with the most conspicuous cases of broad wings always sitting beyond the  $5\sigma$ contour.  One epoch (\#3) sits on the P-Cyg area of the diagram owing to an excess of red-shifted emission and lack of blue-shifted flux. We consider this an \ha\ wind detection, although it does not show the standard P-Cyg shape detected in \he{i}--5876 (Fig. \ref{fig:pcyg}). 

\begin{figure}[t]
\epsscale{1.15}
\plotone{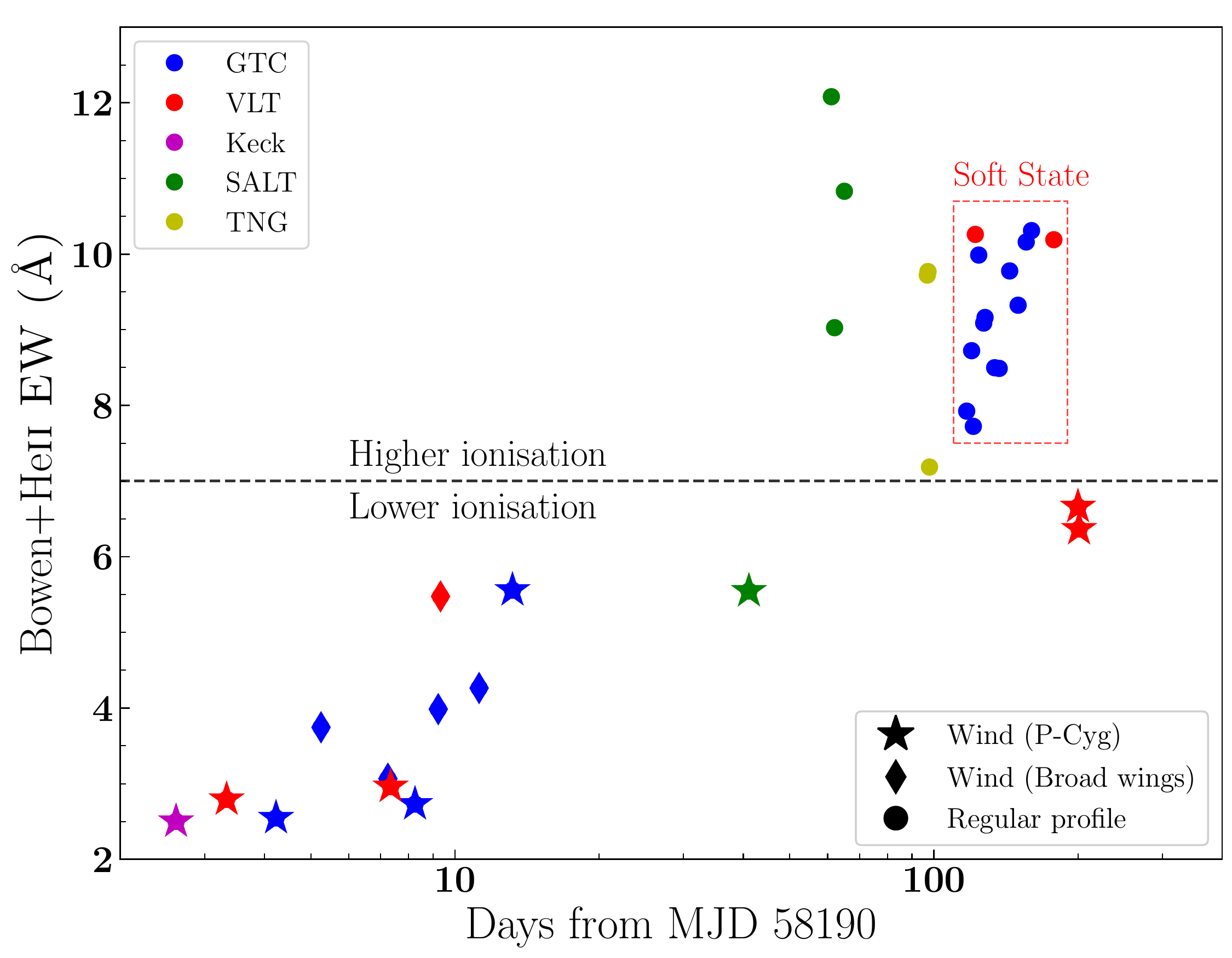}
\caption{Evolution of B+\he{ii} EW with time. Symbols and colors are as in Fig. \ref{fig:hid} (see legend). The horizontal dashed line indicates an empirical threshold in EW above which wind signatures are not detected. \label{fig:bowen}}
\end{figure}

\section{Discussion}
\label{discussion}
We have presented an optical spectroscopic campaign that is one of the most intensive and sensitive ever carried out on a BH transient in outburst. We have detected wind signatures in the form of P-Cyg profiles and broad emission line wings  in several observing epochs during the hard states of the outburst of J1820+070. Such features have been observed before in the BH transients V404 Cyg and V4641 Sgr, although they undergo non-standard outbursts, when only the X-ray hard state, accompanied by strong radio flaring and high X-ray absorption is observed. J1820+070, on the other hand, displayed a standard outburst, showing both hard and soft states across the 8 months spanned by our observations (\citealt{Shidatsu2019}; Fig. \ref{fig:hid}). It is important to note that the wind signatures seen in J1820+070 are significantly shallower than observed in the other two systems (e.g. 1--2 per cent below the continuum level vs. $\sim 20$ per cent). Their detection has only been possible thanks to the combination of the exceptional brightness of the source and our systematic monitoring at very high signal-to-noise. Therefore, these could have been easily missed in other systems, suggesting that  cold winds could be a common feature of black hole accretion in X-ray binaries.

X-ray winds are typically observed during soft states, but are rarely ever present during BH hard states. Hence, these are mostly detected when the radio jet is quenched \citep{Neilsen2009, Ponti2012}, albeit there are some counter-examples (e.g. \citealt{King2015}; see also \citealt{Ponti2014,DiazTrigo2016,Homan2016} for neutron star X-ray winds). Fig. \ref{fig:hid} reveals that  the optical wind is almost ubiquitously detected during the hard state, extensively during outburst rise, but also during the decay towards the end of the soft-to-hard transition (Fig. \ref{fig:pcyg}). In Fig. \ref{fig:bowen} we represent the evolution of the EW of the B+\he{ii} emission lines, which -- given their higher ionization potential -- can be used as a tracer for the state of the disk.  All wind detections are found below EW $\sim7$ \AA. This empirical threshold, which is not meant to be accurate and it is also sensitive to variations in the underlying continuum, excludes the whole soft state and some (typically the softest) hard state epochs (Fig. \ref{fig:hid}). A possible exception to this could be the first soft state epoch, where a shallow blue-shifted absorption might be present in \he{i}--5876 ($\#19$; Fig. \ref{fig:pcyg}). The  B+\he{ii} EW from this epoch is just above the $\sim7$ \AA\ threshold. Therefore, Fig. \ref{fig:bowen} suggests that the non-detection of optical wind features during the soft state might be related to an over-ionisation of the ejecta. This was shown to be a key factor for the detectability of the outflow in V404 Cyg (\citetalias{Munoz-Darias2016}). In addition, some of our most conspicuous wind detections occur at the hard state peak (epochs $\#5$ to $\#11$; blue rectangle in Fig. \ref{fig:hid}), a stage when radio emission is present (Bright et al. in prep.) and strong jet activity is typically witnessed in BH transients \citep[e.g.][]{Fender2004}. This is also the peak of the optical outburst ($g\sim 11.2$; about 0.8 mag brighter than the soft state peak), a fact that has been interpreted as being the result of significant jet contribution to the optical regime \citep{Shidatsu2018, Shidatsu2019}. Therefore, it is clear that the optical wind of J1820+070 is simultaneous with the jet.  

We have measured two characteristic wind velocities: \vt $\sim1200$ and $\sim1800$ \kms. Sometimes these are simultaneously observed in several features of the same spectrum ($\# 7$), but also in observations taken months apart. This suggests a rather complex wind structure, and a very appealing scenario would be that they represent different wind-launching mechanisms.   However, more detailed modelling, beyond the scope of this discovery work, is needed to support such a scenario (see \citealt{Munoz-Darias2018} for a discussion on optical wind launching mechanisms).
In any case, the obtained \vt\ are similar to outflow velocities derived from (hard state) optical winds and (soft state) X-ray winds in other systems (e.g. \citetalias{Munoz-Darias2016}, \citealt{Ponti2016}). This raises the question of whether or not hot and cold winds are intrinsically different or just different observables of the same outflow at different stages of the outburst. In this regard, photoionisation instability curves computed by \citet{Bianchi2017} show that the highly ionised soft state wind might become either fully ionised (hot and low density) or almost neutral (cold and dense) once the system moves to the hard state (and the irradiating spectrum changes accordingly). We are not aware of any X-ray wind detection in J1820+070\footnote{Preliminary analysis of NICER data does not show evidence for soft state X-ray winds (J. Homan private communication)}, but we note that inclination effects play a key role in the detectability of hot outflows \citep{Ponti2012}. Nevertheless, the detection of X-ray dips during the hard state (\citealt{Homan2018,Kajava2019}) and the shape of the hardness-intensity diagram (e.g. pronounced hard state peak; \citealt{Munoz-Darias2013b}) advocates for a relatively high inclination, which would favour the detection of X-ray outflows.

Finally, it is worth mentioning that the shallowness of the P-Cyg features that we have discovered in J1820+070 might be indicative of a lower mass outflow rate \citep[e.g.][]{Castor1979} than for the cases of V404 Cyg and V4641 Sgr, where winds have been proposed to significantly affect the outburst evolution (e.g. \citetalias{Munoz-Darias2016}). In particular, an outflow mass of  $\sim$ 100 times the accreted mass has been estimated for V404 Cyg (\citealt{Casares2019}). This could explain the standard outburst displayed by J1820+070 as compared to these objects. Nevertheless, we note that mass outflow rates up to $\sim 10$ times the accretion rate have been proposed for systems displaying regular outbursts and X-ray winds (e.g. \citealt{Ponti2012}).

\section{Conclusions}
We have detected optical accretion disk winds during the hard states of the outburst in MAXI~J1820+070. Wind signatures are not detected in the soft state, but the visibility of the wind during this stage might be affected by a higher ionisation of the accretion disk/ejecta. The detection of the wind has only been possible thanks to the brightness of the source and our exceptionally intensive and sensitive monitoring, implying that similar outflows are likely present in (at least) a significant fraction of BH X-ray binaries. Therefore, wind-like outflows would not be exclusive of bright, hot states but a common mass and angular momentum loss mechanism that operates through most of the accretion episode.

\section{acknowledgements}
We acknowledge support by the Spanish MINECO under grant AYA2017-83216-P. TMD and MAPT acknowledge support via Ram\'on y Cajal Fellowships RYC-2015-18148 and RYC-2015-17854. TMD is thankful to Alex Fullerton for useful discussion on winds from massive stars. DMS acknowledges support from the ERC under the European Union’s Horizon 2020 research and innovation programme (grant agreement No. 715051; Spiders). PAC is grateful to the Leverhulme Trust for the award of an Emeritus Fellowship. JJEK acknowledges support from the Academy of Finland grant 295114. Some of these observations were obtained with the Southern African Large Telescope under the Large Science Programme on transients, 2016-2-LSP-001 (PI: DAHB). Polish support of this SALT programme is funded by grant no. MNiSW DIR/WK/2016/07. DAHB, EJK and JT acknowledge research support from the South African National Research Foundation. DS acknowledged support from STFC via grant ST/P000495/1. {\sc molly} software developed by Tom Marsh is gratefully acknowledged. Based on observations collected at ESO under programmes 0100.D-0292(A) and 0101.D-0158(A).
\bibliographystyle{aasjournal}
\bibliography{/Users/tmd/Dropbox/Libreria} 



\end{document}